\documentstyle[preprint,aps,epsfig]{revtex}
\tighten

\begin{document}

\preprint{\parbox[t]{50mm}
{\begin{flushright}
ADP-98-70/T337\\
November 1998
\end{flushright}}}

\draft
%
\title{Chiral symmetry breaking in dimensionally regularized\\
 nonperturbative quenched QED }
\author{
   V.~P.~Gusynin,$^{1,2}$ A. W.\ Schreiber,$^{2,3}$
T. Sizer $^{2,3}$ and A.~G.\ Williams $^{2,3}$ }

\address{\phantom{}$^1$Bogolyubov Institute for Theoretical Physics,
  Kiev, 252143 Ukraine\\
   \phantom{}$^2$Special Research Centre for the Subatomic
Structure of Matter, University of Adelaide, 5005, Australia\\
         \phantom{}$^3$Department of Physics and Mathematical Physics,
                   University of Adelaide, 5005, Australia }
%
\maketitle
%
\begin{abstract}

In this paper we study dynamical chiral symmetry breaking in 
dimensionally regularized quenched QED within the context of Dyson-Schwinger
equations.  In $D < 4$ dimensions the theory has solutions which exhibit
 chiral symmetry breaking
for all values of the coupling.  To begin with, we
study this phenomenon both numerically and, with some approximations,
analytically within the rainbow approximation in the Landau gauge.  In
particular, we discuss how to extract the critical coupling
$\alpha_c={\pi \over 3}$ relevant in $4$ dimensions from the $D$
dimensional theory. We further present analytic results for the
chirally symmetric solution obtained with the Curtis-Pennington vertex as 
well as numerical results
for solutions exhibiting chiral symmetry breaking.  For these we demonstrate 
that,
using dimensional regularization,
the extraction of the critical coupling relevant for this vertex is
feasible.  Initial results for this critical coupling are in agreement with
cut-off based work within the currently achievable numerical precision.

\end{abstract}

\section{Introduction}
\label{sec_intro}

It is fairly well established that quantum electrodynamics (QED), and
in particular quenched QED, breaks chiral symmetry for sufficiently
large couplings.  This phenomenon has been observed both in 
lattice simulations~\cite{Lat1} as well as
various studies based on the use of Dyson-Schwinger
equations~\cite{FGMS,MiranskReview,TheReview}. These latter
calculations have generally relied on the use of a cut-off in
euclidean momentum in order to regulate divergent integrals, a
procedure which breaks the gauge invariance of the theory.

On the other hand, continuation of gauge theories to $D < 4$
dimensions has long been used as an efficient way to regularize
perturbation theory without violating gauge invariance.  In
nonperturbative calculations, however, the use of this method of
regularization is rarely used~\cite{eff_the}.  Within the context of the
Dyson-Schwinger equations (DSEs) only a few
publications~\cite{dim_reg,dim_reg_2} have employed dimensional
regularization instead of the usual momentum cut-off.

It is the purpose of the present paper to study dynamical chiral
symmetry breaking and the chiral limit within dimensionally
regularized quenched QED.  We are motivated to do this by the wish to avoid
some gauge ambiguities occurring in cut-off based work, which we
discuss in Sec.~\ref{sec_motivation}. In that Section we also outline
some general results which one expects to be valid for $D < 4$,
independently of the particular vertex which one uses as an input to
the DSEs.  Having done this we proceed, in Section~\ref{sec:
rainbow}, with a study of chiral symmetry breaking in the popular, but
gauge non-covariant, rainbow approximation. Just as in cut-off
regularized work, the rainbow approximation provides a very good
qualitative guide to what to expect for more realistic vertices and
has the considerable advantage that, with certain additional
approximations, one may obtain analytical results.  We  check
numerically that the additional approximations made are in fact quite
justified.  Indeed, it is very fortunate that it is possible to obtain
this analytic insight into the pattern of chiral symmetry breaking in
$D$ dimensions as it provides us with a well defined procedure for
extracting the critical coupling of the 4 dimensional theory with
more complicated vertices.  We proceed to the Curtis-Pennington (CP)
vertex in Section~\ref{sec: cp}.  There we derive, for solutions which do
not break chiral symmetry, an integral representation for the
exact wavefunction renormalization function ${\cal Z}$ in $D$ dimensions.  
We also provide an approximate, but explicit, expression for this quantity.
The latter is quite useful, in the ultraviolet region, even if dynamical chiral symmetry 
breaking takes place as it provides a welcome
check for the numerical investigation of chiral symmetry breaking with
the CP vertex with which we conclude that section.  Finally, in
Section~\ref{sec: conclusion}, we summarize our results and conclude.

\section{Motivation and general considerations}
\label{sec_motivation}

Although chiral symmetry breaking appears to be
universally observed independently of the precise nature of the vertex used
in DSE studies, it has also been recognized for a long time that the
critical couplings with almost all\footnote{Some vertex Ans\"atze
exist which lead to critical couplings which are strictly gauge
independent~\protect\cite{Kondo,BP1,BP2}.  However, these involve
either vertices which have unphysical singularities or ensure 
gauge independence of the critical coupling by explicit construction.} of
these vertices show a gauge dependence which  should not be
present for a physical quantity.  With a bare vertex this is not
surprising as this vertex Ansatz breaks the Ward-Takahashi
identity.  However, even with the Curtis-Pennington (CP) vertex,
which does not violate this identity and additionally is constrained
by the requirement of perturbative multiplicative renormalizability, a
residual gauge dependence remains~\cite{CPIV,ABGPR}.

Apart from possible deficiencies of the vertex, which we do not
investigate in this paper, the use of cut-off regularization
explicitly breaks the gauge symmetry even as the cut-off is taken to
infinity.  This is well known in perturbation theory (see, for example, the
discussion of the axial anomaly in Sect.~19.2 of Ref.~\cite{peskin})
 and was pointed out by Roberts and
collaborators~\cite{dongroberts} in the present context.  The latter
authors proposed a prescription for dealing with this ambiguity which
ensures that the regularization does not violate the Ward-Takahashi identity.

As may be observed in Fig.~\ref{fig: cut-off alpha_cr},
this ambiguity has a strong effect on the value of the critical coupling
of the theory.  The two curves in  that figure correspond to the
critical coupling $\alpha_c$ of Ref.~\cite{ABGPR} as well as the coupling
$\alpha_c'$ one obtains by following the prescription of Roberts et al.
It is straightforward to show, following the analysis of Ref.~\cite{ABGPR},
that these couplings are related through
\begin{equation}
\alpha_{c}' \> = \> { \alpha_{c} \over 1 +
{\xi \> \alpha_{c}  \over 8 \pi}}\;\;\;.
\end{equation}
Also plotted in this figure are previously published numerical 
results~\cite{CPIV,qed4_hrw} obtained with both of the above prescriptions.
Note that, curiously, the critical couplings obtained with the
prescription of Ref.~\cite{dongroberts} (i.e. the calculation which restores
the Ward-Takahashi identity) exhibits a stronger gauge
dependence, at least for the range of gauge parameters shown in 
Fig.~\ref{fig: cut-off alpha_cr}.

Gauge ambiguities such as the one outlined above are absent if one does not
break the gauge invariance of the theory through the regularization
procedure.  Hence, we now turn to dimensionally regularized (quenched)
QED. The Minkowski space fermion propagator $S(p)$ is defined in the usual 
way through
the dimensionless wavefunction renormalization function ${\cal Z}(p^2)$
and the dimensionful mass function $M(p^2)$, i.e.
\begin{equation}
S(p) \> \equiv \> {{\cal Z}(p^2) \over {p \hspace{-5pt}/} - M(p^2)}\;\;\;.
\end{equation}
The dependence of ${\cal Z}$ and $M$
on the dimensionality of the space is not explicitly
indicated here.  Furthermore, note that 
to a large extent we shall be dealing only 
with the  regularized theory without imposing a renormalization procedure, as  
renormalization~\cite{qed4_hrw,renorm} is inessential to our discussion. 

In addition to the above, we shall consider the theory without explicit chiral 
symmetry
breaking (i.e. zero bare mass).  This theory would not contain a mass
scale were it not for the usual arbitrary scale (which we
denote by $\nu$) introduced in $D=4 - 2 \epsilon$ dimensions
which provides the connection between
the {\it dimensionful} coupling $\alpha_D$ and the usual
{\it dimensionless} coupling
constant $\alpha=e^2/4\pi$:
\begin{equation}
\alpha_D \> = \> \alpha \> \nu^{2 \epsilon}\;\;\;.
\end{equation}
As $\nu$ is {\it the only} mass scale in
the problem, and as the coupling always appears in the above combination with
this scale, on dimensional grounds alone the mass function must be of the form
\begin{equation}
M(p^2) \> = \> \nu \>  \alpha^{1 \over 2 \epsilon} \> \tilde M
\left( {p^2 \over \nu^2 \alpha^{1 \over \epsilon} },\epsilon \right)
\end{equation}
where $\tilde M$ is a dimensionless function and in
particular
\begin{equation}
M(0) \> = \> \nu \>  \alpha^{1 \over 2 \epsilon} \>
\tilde M \left(0,\epsilon \right)\;\;\;.
\label{eq: M(0) general form}
\end{equation}
Moreover, as $\epsilon$ goes to zero the $\nu$ dependence on the right hand
side must disappear and hence the dynamical mass $M(0)$ is either
zero (i.e. no symmetry breaking) or goes to infinity in this limit.
This situation is analogous to what happens in cut-off regularized theory,
where the scale parameter is the cut-off itself and the mass is proportional
it.

Note that $\tilde M(0,\epsilon)$ is not dependent on $\alpha$. This
implies immediately that there can be no non-zero critical coupling
in $D \ne 4$ dimensions: if $M(0)$ is non-zero for some coupling
$\alpha$ then it must be non-zero for all couplings.

Given these general considerations (which are of course independent of
the particular Ansatz for the vertex) it behooves one to ask how this
situation can be reconciled with a critical coupling $\alpha_c$ of order 1 in
four dimensions.  In order to see how this might arise, we shall extract
a convenient numerical factor out of $\tilde M$ and suggestively re-write
the dynamical mass as
\begin{equation}
M(0) \> = \> \nu \>  \left({\alpha \over \alpha_c}\right)^{1 \over 2 \epsilon} \>
\overline M \left(0,\epsilon \right)\;\;\;.
\label{eq: M(0) general form 2}
\end{equation}
At present there is no difference in content between
Eq.~(\ref{eq: M(0) general form}) and Eq.~(\ref{eq: M(0) general form 2}).
However, if we now {\it define} $\alpha_c$ by demanding that
the behaviour of $M(0)$ is
dominated by the factor $\left({\alpha \over \alpha_c}\right)^{1/2\epsilon}$
as $\epsilon$ goes to zero, which is equivalent to demanding that
\begin{equation}
[\overline M \left(0,\epsilon \right)]^\epsilon \>
\longrightarrow_{_{_{_{\hspace{-7mm}
{{ \epsilon\rightarrow}{ 0}}}}}}
\>\> 1 \;\;\;,
\label{eq: m_overbar def}
\end{equation}
then the intent becomes clear: even though $M(0)$ may be nonzero for
all couplings in $D < 4$ dimensions, in the limit that $\epsilon $ goes
to zero we obtain
\begin{eqnarray}
M(0)\>
\longrightarrow_{_{_{_{\hspace{-7mm}
{{ \epsilon\rightarrow}{ 0}}}}}}
\> \> 0
&& \hspace{2cm} \alpha \> < \alpha_c \\ \nonumber
M(0)\>
\longrightarrow_{_{_{_{\hspace{-7mm}
{{ \epsilon\rightarrow}{ 0}}}}}}\> \>\infty
&& \hspace{2cm} \alpha \> > \alpha_c \;\;\;. \nonumber
\end{eqnarray}

Note that in the above we have not addressed the issue of
whether or not there actually is a
$\overline M\left(0,\epsilon \right)$ with the property
of Eq.~(\ref{eq: m_overbar def}).  In fact, the numerical
and analytical work in the following sections is
largely concerned with finding this function and hence determining
whether or not chiral symmetry is indeed broken for
$D < 4$.\footnote{The reader will note that  as neither
$\tilde M(0,\epsilon)$ nor $\overline M\left(0,\epsilon \right)$ are functions
of the coupling $\alpha$, the value of $\alpha_c$ can be determined independently
of the strength $\alpha$ of the self-interactions in $D<4$ dimensions.}
Notwithstanding this, as one knows from cut-off based work
that there actually is a non-zero
critical coupling for $D=4$, one can at this stage already come
to the conclusion that  $\overline M\left(0,\epsilon \right)$ exists and
hence that quenched QED in $D < 4$ dimensions has a chiral symmetry 
breaking solution for all couplings.  

In summary, as the trivial solution $M(p^2)=0$ always exists as well,
we see that in $D<4$ dimensions the trivial and symmetry breaking solutions
bifurcate at $\alpha=0$ while for $D=4$ the point of bifurcation is at
$\alpha=\alpha_c$; i.e., there is a discontinuous change in the
point of bifurcation. As $D$ approaches four (i.e. as $\epsilon$ approaches $0$)
 the generated mass $M(0)$ decreases (grows) roughly like $\left (\alpha/
\alpha_c \right )^{1/2\epsilon}$ for $\alpha$ $\lesssim$ $\alpha_c$ 
($\gtrsim$ $\alpha_c$),
respectively, becoming an infinite step function at $\alpha=\alpha_c$ 
when $\epsilon$
goes to zero.

\section{The rainbow approximation}
\label{sec: rainbow}
Let us now consider an explicit vertex.
To begin with, we consider the rainbow approximation to the
Euclidean mass function of quenched QED with zero bare mass
in Landau gauge.  It is given by
\begin{equation}
M(p^2)=(e \nu^{\epsilon})^2(3 - 2 \epsilon)\int \;\frac{d^Dk}{(2\pi)^D}\;
\frac{M(k^2)}
{k^2+M^2(k^2)}\frac{1}{(p-k)^2}\;\;\;.
\label{masseq}
\end{equation}
Note that
the Dirac part of the self-energy is equal to zero in the Landau gauge
in rainbow approximation even in $D < 4$ dimensions and hence that
${\cal Z}(p^2)=1$ for all $p^2$.

It is of course possible to find the solution to Eq.~(\ref{masseq})
numerically -- indeed we shall do so -- however it is far more instructive
to first try to make some reasonable approximations in order to be able
to analyze it analytically.
First, as the angular integrals involved
in $D$-dimensional integration are standard (see, for example,
Refs.~\cite{dim_reg_2} and ~\cite{Muta}) we may reduce Eq.~(\ref{masseq})
to a one-dimensional integral, namely
\begin{eqnarray}
M(p^2)&=&\alpha \nu^{2 \epsilon}c_{\epsilon} \>\int
\limits_0^\infty \frac{dk^2 (k^2)^
{1-\epsilon}M(k^2)}{k^2+M^2(k^2)}\left[{1\over p^2}F\left(1,\epsilon;
2 - \epsilon;\frac{k^2}{p^2}\right)\theta(p^2-k^2)\right.\nonumber\\
&+&\left.{1\over k^2}F\left(1,\epsilon;2-\epsilon;\frac{p^2}{k^2}\right)
\theta(k^2-p^2)\right],
\label{inteq}
\end{eqnarray}
where
\begin{equation}
c_{\epsilon}= \frac{3 - 2 \epsilon}{(4\pi)^{1-\epsilon}\Gamma(2-\epsilon)},
\quad (c_0=\frac{3}{4\pi}).
\end{equation}
Note that for $D=4$ the mass function in Eq.(\ref{inteq}) reduces to the standard one in
QED$_4$.

In $D\ne 4$ dimensions the hypergeometric functions in Eq.~(\ref{inteq})
preclude a solution in closed form.  However, note that these
hypergeometric functions have a power expansion in $\epsilon$ so that
for small $\epsilon$ one is not likely to go too far wrong by just replacing
these by their $\epsilon = 0$ (i.e. $D=4$) limit.  After all, the reason
for choosing dimensional regularization in the first place is in order
to regulate the integral, and this is achieved by the  factor of
$k^{- 2 \epsilon}$, not the hypergeometric functions.  In addition, this
approximation also corresponds to just replacing the hypergeometric functions
by their IR and UV limits, so that one might expect that even for larger
$\epsilon$ that the approximation is not too bad in these regions
\footnote{It is however possible to show
that a linearized version of Eq.~(\ref{masseq}) always
has symmetry breaking solutions even without making this approximation
of the angular integrals.  We indicate how this may be done in 
Appendix A.}.

Making this replacement, i.e.
\begin{equation}
M(p^2)=\alpha  \nu^{2 \epsilon}c_{\epsilon}\left[{1\over p^2}\int\limits_0^{p^2}\frac{dk^2 \>
(k^2)^{1 - \epsilon} M(k^2)}{k^2+M^2(k^2)}+\int\limits_{p^2}^\infty\frac{dk^2
\>(k^2)^{-\epsilon} M(k^2)}{k^2+M^2(k^2)}\right]\;\;\;,
\label{inteqxa}
\end{equation}
allows us to convert Eq.~(\ref{inteq}) into a differential equation, namely
\begin{equation}
\left[p^4 M^\prime(p^2)\right]^\prime+\alpha  \nu^{2 \epsilon}c_{\epsilon}\frac{(p^2)^{1 -
\epsilon}}{p^2+M^2(p^2)}M(p^2)=0\;\;\;,
\label{diffeq}
\end{equation}
with the boundary conditions
\begin{equation} p^4 M^\prime(p^2)|_{p^2=0}=0,\qquad \left[p^2 M(p^2)\right]^\prime|_{p^2=
\infty}=0\;\;\;.
\end{equation}
Unfortunately, the differential equation (\ref{diffeq}) still has no
solutions in terms of known special functions. Since the mass function in the
denominator of Eq. (\ref{inteqxa}) serves primarily as an infrared regulator
we shall make one last approximation and replace it  by an infrared
cut-off for the integral, which can be taken as a fixed value of
$M^2(k^2)$ in the infrared region (for convenience we shall call this value
the `dynamical mass' $m$).  This simplifies the problem sufficiently to allow
the derivation of an analytical solution.

In terms of the dimensionless variables $x=p^2/\nu^2$, $y=k^2/\nu^2$ and $a=m^2/
\nu^2$ the linearized equation becomes
\begin{equation}
M(x)=\alpha c_{\epsilon}\left[{1\over x}\int\limits_a^x\frac{dy \> y^{1 - \epsilon}
M(y)}{y}+\int\limits_x^\infty\frac{dy \>y^{-\epsilon} M(y)}{y}\right]\;\;\;;
\label{inteqx}
\end{equation}
for simplicity, we do not explicitly 
differentiate between $M(x)$ and $M(p^2)$.
This may be written in differential form as
\begin{equation}
\left[x^2 M^\prime(x)\right]^\prime\> +\> \alpha \>c_{\epsilon} \>x^{-\epsilon} M(x)=0
\;\;\;,
\label{diffeqsimple}
\end{equation}
with the boundary conditions
\begin{equation}
M^\prime(x)|_{x=a}=0,\qquad \left[x M(x)\right]^\prime|_{x=\infty}=0\;\;\;.
\label{BC}
\end{equation}
This differential equation  has solutions in terms of Bessel
functions
\begin{equation}
M(x)=x^{-1/2}\left[C_1\> J_{\lambda}
\left(\frac{\sqrt{4\alpha c_{\epsilon}}}{\epsilon \>  x^{\epsilon \over 2}}\right)
+C_2 \> J_{-{\lambda}}
\left(\frac{\sqrt{4\alpha c_{\epsilon}}}{\epsilon \>  x^{\epsilon \over 2}}\right)\right]\;\;\;,
\label{diffeqsol}
\end{equation}
where we have defined $\lambda = 1/\epsilon$ in order to avoid
cumbersome indices on the Bessel functions.
The ultraviolet boundary condition Eq.~(\ref{BC}) gives $C_2=0$ while
the infrared boundary condition leads to 
\begin{equation}
C_1 \> \left[J_{\lambda}\left(\frac{\sqrt{4\alpha c_{\epsilon}}}{\epsilon\>  x^{\epsilon \over 2}}\right)+
\frac{\sqrt{4\alpha c_{\epsilon}}}{x^{\epsilon \over 2}}
J_{\lambda}^\prime\left(\frac{\sqrt
{4\alpha c_{\epsilon}}}{\epsilon\>  
x^{\epsilon \over 2}}\right)\right]_{x=a}\> =\> 0\;\;\;.
\label{dynmasseq}
\end{equation}
This equation may be simplified
using the relation among Bessel functions
\begin{equation}
zJ_\lambda^\prime (z)+\lambda J_\lambda(z)=zJ_{\lambda-1}(z)\;\;\;,
\end{equation}
and becomes
\begin{equation}
C_1 \> \left[\frac{\sqrt{4\alpha c_{\epsilon}}}{\epsilon \> x^{\epsilon \over 2}}
J_{{\lambda}-1}\left(\frac{\sqrt
{4\alpha c_{\epsilon}}}{\epsilon \> x^{\epsilon \over 2}}\right)\right]_{x=a}\> =0\> \;\;\;.
\label{eq: simple dynmasseq}
\end{equation}
Clearly this equation is satisfied by $C_1=0$, which corresponds to
the trivial chirally symmetric solution $M(x)=0$.  However, for values of $a$ which are
such that the argument of the Bessel function in Eq.~(\ref{eq: simple
dynmasseq}) corresponds to one of its zeroes, the equation is also
satisfied for $C_1 \neq 0$, i.e. for these values of $a$ there exist
solutions with dynamically broken chiral symmetry.  If we define
$j_{\lambda-1,1}\> = \> \sqrt{4\alpha c_{\epsilon}} /\epsilon 
a^{\epsilon/2}$ to be the smallest positive zero of
Eq.~(\ref{eq: simple dynmasseq}), the dynamical mass for this solution
becomes
\begin{equation}
m\> =\> \nu \> a^{1/2}\> =\> \nu \> \alpha^{1 \over
2 \epsilon} \> \left(\frac{\sqrt{4 c_{\epsilon}}}{\epsilon\>
j_{{\lambda}-1,1}}\right)^{1 \over \epsilon}.
\label{dynmass}
\end{equation}
Note that for this solution the normalization $C_1$ is not fixed by
Eq.~(\ref{inteqx}) as this equation is linear in $M(x)$. Later on we
shall fix $C_1$ by demanding that $M(a) = m$, however there is no
 compelling reason to do this and one might alternatively fix the
normalization in such a way as to approximate the true (numerical)
solutions of Eq.~(\ref{masseq}) as well as possible.  Finally, note
that, as expected, a dynamical symmetry breaking solution exists
for any value
of the coupling and that the expression for the dynamical mass is in
agreement with the general form expected from dimensional
considerations [i.e. Eq.~(\ref{eq: M(0) general form})].

In order to extract $\alpha_c$, we need to look at the behaviour of $m$
as $\epsilon$ goes to zero (i.e. $\lambda \rightarrow \infty$). This may
be done by noting that the positive roots of the Bessel function
$J_\lambda$ have the following asymptotic behaviour (see, for example,
Eq.~9.5.22 in Ref.~\cite{abramowitz}):
\begin{equation}
j_{\lambda,s}\sim\lambda
z(\zeta)+\sum\limits_{k=1}^\infty\frac{f_k(\zeta)}{\lambda^{2k-1}},\quad
\zeta=\lambda^{-2/3}a_s,
\end{equation}
where $a_s$ is the $s$th negative zero of Airy function Ai($z$),
and $z(\zeta)$ is determined $(z(\zeta)>1)$ from the equation
\begin{equation}
{2\over3}(-\zeta)^{3/2}=\sqrt{z^2-1}-\arccos{1\over z}.
\end{equation}
For large $\lambda$ the variable $\zeta$ is small and so
it is valid to expand $z$ around 1.  Writing $z=1+\delta$ we obtain
\begin{equation}
\arccos\frac{1}{1+\delta}\sim\sqrt{2\delta}-\frac{5\sqrt2}{12}
\delta^{3/2},
\end{equation}
and so $\delta\simeq -\zeta/{2^{1/3}}$ yielding, to leading order,
\begin{equation}
z=1-\frac{a_1}{2^{1/3}\lambda^{2/3}}\;\;\;.
\end{equation}
If we define
\begin{equation}
\gamma = -{a_1 \over 2^{1/3}} \> \sim \> 1.855757
\end{equation}
then the leading terms in the expansion of $j_{\lambda-1,1}$ are
\begin{equation}
j_{\lambda-1,1}\sim\lambda +\gamma\lambda^{1/3}-1+O(\lambda^{-1/3}).
\label{eq: root}
\end{equation}
Also,  the coefficient $c_{\epsilon}$ appearing in Eq.~(\ref{dynmass}) may be
expanded
\begin{equation}
c_{\epsilon}\hspace{3mm}\sim_{_{_{_{\hspace{-5mm}
{{\epsilon\to 0}}}}}}\frac{3}{4\pi}(1+d \> \epsilon),\quad
d=\ln(4\pi)+{1\over3}+\psi(1)
\end{equation}
so that for small $\epsilon$ the dynamical mass becomes
\begin{equation}
m \> \sim \> \nu \> \alpha^{1 \over 2 \epsilon}
\> {\left[{3 \over \pi} (1+d \> \epsilon) \right]^{1 \over 2 \epsilon} \over
(1+\gamma\epsilon^{2/3}-\epsilon)^{1 \over \epsilon}}
\> \sim \> \nu \> \left(  {\alpha \over {\pi/ 3}}\right)^{1 \over 2 \epsilon}
\> e^{1 + {d \over 2} - \gamma\epsilon^{-1/3}}\;\;\;.
\label{eq: dyn mass app}
\end{equation}
Note that the behaviour of the first term (for $\epsilon$ going to
zero)  dominates over the exponential function, as required
in Eq.~(\ref{eq: m_overbar def}).  Hence the critical coupling
in four dimensions is given by $\pi / 3$, as expected
from cut-off based work~\cite{FGMS,MiranskReview}.

Returning now to the mass function itself,
we may substitute the expression for the dynamical mass, i.e., Eq.~(\ref{dynmass}),
together with our choice of normalization
condition
\begin{equation}
M(p^2=m^2) \> = \> m\;\;\;,
\label{eq: norm cond}
\end{equation}
into Eq.~(\ref{masseq}) in order to eliminate $C_1$.  One obtains
\begin{equation}
M(p)=\frac{m^2}{|p|}\frac{J_\lambda[j_{\lambda-1,1}\cdot\left({m\over
|p|}\right)^{\epsilon}]}{J_\lambda[j_{\lambda-1,1}]}\;\;\;.
\label{massfunction}
\end{equation}
Note that the explicit dependence on $\nu$ (and hence $\alpha$) has been completely
replaced by $m$ in this expression.

So far we have taken $\alpha$ independent of the regularization. As we have seen this
leads to a dynamically generated mass which becomes infinite as the regulator is
removed.  Fomin et al.~\cite{FGMS} examined (within cut-off regularized
QED)  a different limit, namely one where the mass $m$ is kept
constant while the cut-off is removed. 
In our case this limit necessitates that the coupling $\alpha$ is dependent on 
$\epsilon$ through
\begin{equation}
\alpha \simeq {\pi \over 3} \left ( 1 \> + \> 2 \>\gamma \>\epsilon^{2 \over 3} \right )
\end{equation}
[see Eq.~(\ref{eq: dyn mass app}); note that $\alpha_c$ is approached from above].
  The limit may be taken analytically in Eq.~(\ref{massfunction})
by making use of the
known asymptotic behaviour of the Bessel functions (see Eq. 9.3.23 of
Ref.~\cite{abramowitz}), i.e.
\begin{equation}
J_\lambda\left(\lambda+\lambda^{1/3}z\right)\sim\left(\frac{2}{\lambda}\right)^{1/3}
{\rm Ai}(-2^{1/3}z)\;\;\;,
\end{equation}
as well as the asymptotic expansion of $j_{\lambda-1,1}$ in Eq.~(\ref{eq: root}).
One obtains
\begin{equation}
M(p)=\frac{m^2}{p}\left(\ln{p\over m}+1\right)\;\;\;,
\end{equation}
which agrees with the result in Ref.~\cite{FGMS}.

\begin{figure}[htb]
  \centering
      \epsfig{angle=90,height=8cm,file=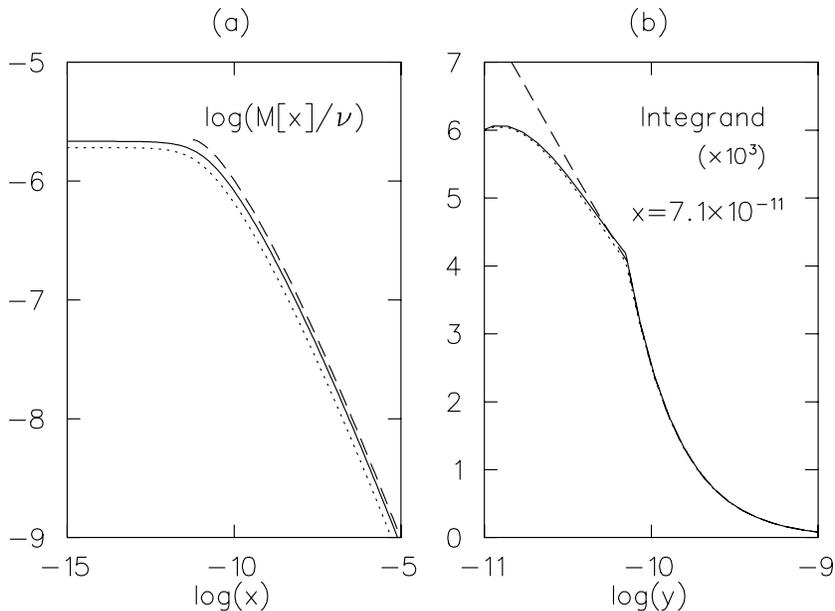}
  \vspace{0.5cm}
  \parbox{130mm}{\caption{ The mass function for rainbow QED for $\alpha=0.6$ and
$\epsilon=0.03$ as a function of $x \equiv p^2/\nu^2$.  The dynamical mass is 
specified by $m/\nu = 2.24 \times 10^{-6} $, where
$\nu$ is the scale introduced by dimensional regularization.
The solid line corresponds to the exact numerical solution of 
Eq.~(\protect\ref{masseq}), the dashed line is the Bessel function
of Eq.~(\protect\ref{massfunction}) and the dotted line is the solution of the
Dyson-Schwinger equation with the hypergeometric function replaced by unity 
(Eq.~\protect\ref{inteqxa}).  In (a) the actual mass function is shown, while in
(b) we show the integrand at a particular value of $x$.
  \label{fig: comp}}}
  \vspace{0.5cm}
\end{figure}

To conclude this section, we analyze the validity of the
approximations made by solving Eq.~(\ref{masseq}) numerically and
comparing it to the Bessel function solution in
Eq.~(\ref{massfunction}).  In Fig.~\ref{fig: comp}a we have
plotted the mass function (divided by $\nu$) as a function of the
dimensionless momentum $x$ for a moderately large coupling ($\alpha =
0.6$) and $\epsilon = 0.03$.  The solid curve corresponds to the exact
numerical result [Eq.~(\ref{masseq})] while the dashed line is a plot of
Eq.~(\ref{massfunction}) for these parameters.  As can be seen, the approximation
is not too bad and could actually be made
significantly better  by adopting a different
normalization condition to that in Eq.~(\ref{eq: norm cond}) 
However, no further insight is gained by doing this and
 we shall not pursue it further.

One might naively think that most of the difference between the Bessel
function and the exact numerical solution comes from the linearization of
Eq.~(\ref{masseq}) -- i.e., the approximation made by going from
Eq.~(\ref{inteqxa}) to Eq.~(\ref{inteqx}), as the only approximation
made prior to this is to replace the hypergeometric functions by
unity, which is expected to be good to order $\epsilon$ (i.e. in this
case, 3 \%).  This turns out to be not the case; the dotted curve in
Fig.~\ref{fig: comp}a corresponds to the (numerical) solution of
Eq.~(\ref{inteqxa}).  Not only is the difference to the true solution
essentially an order of magnitude larger than expected (about 30 \% -- note
that Fig.~\ref{fig: comp}a is a log-log plot),
it is actually of opposite sign to the equivalent difference for the
Bessel function.  In other words, the validity of the two
approximations is roughly of the same order of magnitude and they
tend to compensate.

Why are the quantitative differences rather larger than expected?  On
the level of the integrands the approximations are actually quite
good.  In Fig.~\ref{fig: comp}b we show the integrands of
Eqs.~(\ref{masseq}), (\ref{inteqx}) and Eq.~(\ref{inteqxa}) for a
value of $x$ in the infrared ($x \approx 7.1$ $10^{-11}$).  Clearly
the replacement of the hypergeometric functions by unity is indeed an
excellent approximation, as is the linearization performed in
Eq.~(\ref{inteqx}) (except in the infrared, as expected.  Note that
when estimating the contribution to the integral from different $y$
one should take into account that the x-axis in Fig.~\ref{fig:
comp}b is logarithmic).  The real source of the `relatively large'
differences observed for the integrals in Fig.~\ref{fig: comp}a is
the fact that these are integral equations for the function $M(x)$ --
small differences in the integrands do not necessarily guarantee small
differences in $M(x)$.  To illustrate this point, consider a
hypothetical `approximation' to Eq.~(\ref{inteqx}) in which we just
scale the integrands by a constant factor $1+\epsilon$
and ask the question how much this affects the solution $M(x)$.  For
$x=0$ the answer is rather simple: the hypothetical approximation just
corresponds to a rescaling of $\alpha$ by $1+\epsilon$ and as $M(0)$
scales like $\alpha^{1 \over 2 \epsilon}$ we find that the solution
has increased by a factor $(1+\epsilon)^{1 \over 2 \epsilon}$.  In other words,
even in the limit $\epsilon \rightarrow 0$ there remains a remnant
of the `approximation', namely a rescaling of $M(0)$ by a factor $e^{1/2}
\approx 1.6$!

\section{The Curtis-Pennington vertex}
\label{sec: cp}

We shall now leave the rainbow approximation and turn to the CP vertex.
The expressions for the scalar and Dirac
self-energies for this vertex, using dimensional regularization and in an
arbitrary gauge, have
already been given in Ref.~\cite{dim_reg_2}.  Before we discuss
chiral symmetry breaking for this vertex we shall first examine the
chirally symmetric phase.  We remind the reader that in this phase in four dimensions
the wavefunction renormalization has a very simple form for this
vertex~\cite{CPII}, namely
\begin{equation}
{\cal Z}(x,\mu^2)|_{M(x)=0} \> = \> \left( {x \over \mu^2} \right)^{{\xi \alpha \over 4 \pi}}\;\;\;,
\label{eq: pow z}
\end{equation}
where the renormalized Dirac propagator is given by
\begin{equation}
S(p) \> = \> {{\cal Z}(x,\mu^2) \over {p \hspace{-5pt}/} }\;\;\;.
\end{equation}
Here $\xi$ is the gauge parameter and $\mu^2$ is the (dimensionless)
renormalization scale.
This power behaviour of ${\cal Z}(x)$ is in fact demanded by multiplicative
renormalizability~\cite{Brown} as well as gauge covariance~\cite{dongroberts}.
  We shall derive the form of this self-energy
in $D < 4 $ dimensions, which will provide a very useful check on the numerical
results even if $M(x) \ne 0$ as long as $x >> \left ({M(x) \over \nu}
\right )^2$.

\subsection{${\cal Z}(p^2)$ in the chirally symmetric phase}
\label{sec: cp one}

In the chirally symmetric phase, the unrenormalized ${\cal Z}(x)$ corresponding
to the CP vertex in $D$ dimensions
is given by

\begin{equation}
  {\cal Z}(x)\>  =  \> 1 \> + \>
        \frac{\alpha}{4\pi}
        \frac{(4\pi)^\epsilon}{\Gamma(2-\epsilon)} \> \xi \>
        \int_0^\infty dy \>{ y^{-\epsilon} \over x-y} {\cal Z}(y)
\left[ (1-\epsilon)\left(1 - I_1^D\left({y \over x}\right)\right) + {y \over y+x} I_1^D\left({y \over x}\right)\right].
\label{eq: massless a}
\end{equation}
This equation may be obtained from Eq.~(A6) of Ref.~\cite{dim_reg_2} by setting
 $b(y)$ equal to zero in that equation and by using Eq.~(A8) of the same reference
 in order to eliminate the terms with coefficient $a^2(y)$.  The angular integral
 $I_1(w)$ is
defined to be
\begin{eqnarray}
I_{1}^D(w) & = & (1+w) \,\,{}_2F_1(1,\epsilon;2-\epsilon;w)\quad 0 \leq w \leq 1\\
I_{1}^D(w) & = & I_{1}^D (w^{-1} )\quad \quad \quad \quad \quad \quad
\quad \quad \quad w \geq 1\;\;\;.
\end{eqnarray}
In four dimensions the solution to Eq.~(\ref{eq: massless a}) is given
by a ${\cal Z}(x)$ having a simple power behaviour while for $D < 4$ this
is clearly no longer the case.  Nevertheless, it is
possible to derive an integral representation of the solution of Eq.~(\ref{eq: massless a})
by making use of the gauge covariance of this equation.  We do so in 
Appendix B, with the result
\begin{equation}
{\cal Z}(x)
\>=\> x^{\epsilon \over 2} \> 2^{1 - \epsilon}  \> \Gamma(2-\epsilon)
\int_0^\infty  du\>  u^{\epsilon-1} e^{-r u^{2 \epsilon}} J_{2 - \epsilon}({\sqrt x} \> u)
\;\;\;.
\label{eq: res}
\end{equation}
Although this result is exact it is somewhat cumbersome to evaluate numerically because, for
$\epsilon \rightarrow 0$, the oscillations in the integrand become increasingly important.  
For this reason
we shall   
approximate the integrand in Eq.~(\ref{eq: massless a}) by its IR and UV limits, 
as we did for the rainbow approximation
 (as before, this approximation is
good to order $\epsilon$).  Using
\begin{equation}
I_{1}^D(w) \>= \> 1 + {2 \over 2 - \epsilon}w \> + \> O[w^2]
\end{equation}
this approximation yields
\begin{equation}
  {\cal Z}(x)\>  =  \> 1 \> + \>
        \frac{\alpha}{4\pi}
        \frac{(4\pi)^\epsilon}{\Gamma(2-\epsilon)} \> \xi \>
        \left[{\epsilon \over 2 - \epsilon}
         \int_0^x dy \>{ y^{1-\epsilon}\over x^2} {\cal Z}(y) 
         \> - \>  \int_x^\infty dy \> y^{-\epsilon-1} {\cal Z}(y) \right]\;\;\;.
\label{eq: massless z approx}
\end{equation}
This may be converted to the differential equation
\begin{equation}
{\cal Z}''(x) \> + \> {3 \over x} {\cal Z}'(x)
\> = \>         \frac{\tilde c}{x^{1+\epsilon}}  \>
\left[ {\cal Z}'(x) \> + \> 2 \>{1 - \epsilon \over x} \>{\cal Z}(x) \right]
\label{eq: z diff eq}
\end{equation}
where $\tilde c$ is defined to be
\begin{equation}
\tilde c \> = \> \frac{\alpha}{2\pi}
        \frac{(4\pi)^\epsilon}{\Gamma(3-\epsilon)} \> \xi
\end{equation}
and the appropriate boundary conditions are
\begin{equation}
x^{2-\epsilon} {\cal Z}(x)|_{x=0}=0,\qquad {\cal Z}(x)|_{x=\infty}=1.
\end{equation}
(The IR boundary condition arises from the requirement that the
integral in Eq.~(\ref{eq: massless z approx}) needs to converge at its lower limit.)  
In order to solve Eq.~(\ref{eq: z diff eq}), it is convenient
to change variables to
\begin{equation}
z \> = \> {\tilde c \over 2 - \epsilon} \> x^{- \epsilon}\;\;\;,
\end{equation}
and to define
\begin{equation}
a \> = \> {2 \over \epsilon} - 1
\end{equation}
so that the differential equation becomes
\begin{equation}
z {\cal Z}'' \> - \> a (1-z) {\cal Z}'
\> - \> a (a-1) {\cal Z} \> = \> 0\;\;\;,
\end{equation}
while the boundary conditions now are
\begin{equation}
z^{-a} {\cal Z}|_{z=\infty}=0,\qquad {\cal Z}|_{z=0}=1.
\end{equation}

This equation is essentially  Kummer's Equation
(see Eq. 13.1.1 of Ref.~\cite{abramowitz}; we use the notation
of that reference in the following).  Its general solution
may be expressed in terms of confluent hypergeometric functions, i.e.
\begin{eqnarray}
{\cal Z} &=& z^{a+1} e^{-a  z} \> 
\left [
\tilde C_1 \> M(a,a+2; a z) \> + \> \tilde C_2 \> U(a,a+2; a z) 
\right ]\nonumber \\
&=&
e^{-a  z} \>  \left \{C_1 \> \left [
\gamma(a+1, - a z) \> + \> a z \>  \gamma(a,   -a z ) \right ] 
\> + \> C_2 \>[1+z]\right \}\;\;\;.
\label{eq: gen_solution}
\end{eqnarray}
The UV boundary condition is fulfilled if $C_2 = 1$ while $C_1$ is not fixed by
the boundary conditions.  Although Eq.~(\ref{eq: massless z approx}) is solved by
Eq.~(\ref{eq: gen_solution}) for arbitrary $C_1$ we shall concentrate on the
solution with $C_1$=0.  The reason for this is that the solution to the
unapproximated integral [Eq.~(\ref{eq: res})] vanishes at $x=0$ 
(see Appendix B) while the term multiplying $C_1$ in Eq.~(\ref{eq: gen_solution})
diverges like $x^{2 \epsilon - 2}$ and is therefore unlikely to provide a
good approximation to Eq.~(\ref{eq: massless a}).
Hence we obtain
\begin{equation}
{\cal Z}(x) \> = \> \left[1 \> + \> {\tilde c \over 2 - \epsilon}  x^{-\epsilon} \right]
\exp(-{\tilde c \over \epsilon} x^{-\epsilon})\;\;\;.
\end{equation}
Finally, the renormalized function ${\cal Z}(x,\mu^2)$ may be obtained from this
by demanding that  ${\cal Z}(\mu^2,\mu^2) = 1$ so that the renormalized
wavefunction renormalization becomes
\begin{equation}
{\cal Z}(x,\mu^2) \> = \>
{1 \> + \> { \tilde c\over 2 - \epsilon}  x^{- \epsilon}
\over  1 \> + \> {\tilde c \over 2 - \epsilon}  \mu^{-2 \epsilon} }
\>{ \exp(-{\tilde c \over \epsilon} x^{- \epsilon})
\over
\exp(-{\tilde c \over \epsilon} \mu^{-2 \epsilon})}\;\;\;.
\label{eq: cp z analytic}
\end{equation}
Only in the limit $D \rightarrow 4$ does this reduce to the usual power
behaved function found in cut-off based work [Eq.~(\ref{eq: pow z})]
while for $D < 4$ it vanishes non-analytically at $x=0$.  On the other
hand, note that the solution to Eq.~(\ref{eq: massless a}) -- for
finite $\epsilon$ -- only goes to zero linearly in  $x$. For the purpose of 
this paper
this difference in the analytic behaviour in the infrared does not concern us
as for solutions which break chiral symmetry the infrared region is regulated 
by 
$M^2(x)$ so that we do not expect the chirally symmetric ${\cal Z}$ to be
a good approximation in this region in any case.

\subsection{Chiral symmetric breaking for the CP vertex}
\label{sec: cp two}

\begin{figure}[htb]
  \centering
      \epsfig{angle=90,height=7.5cm,file=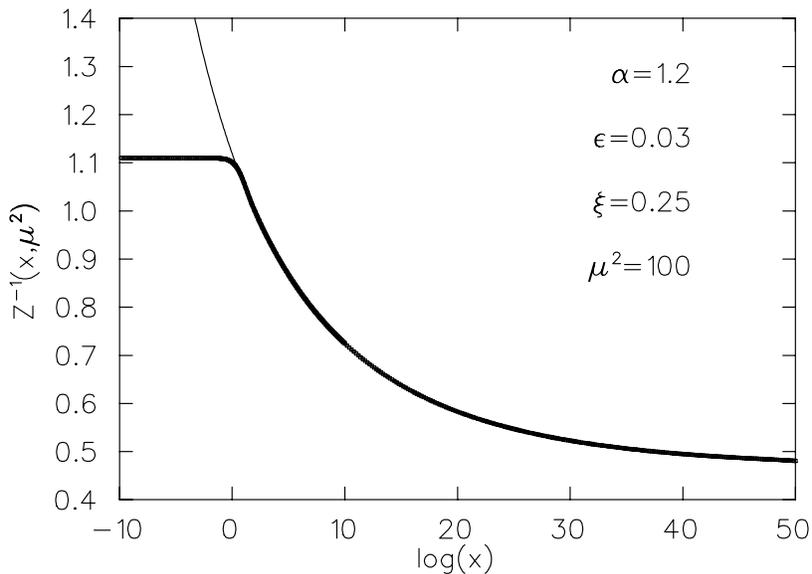}
  \parbox{130mm}{\caption{A typical (inverse) wave function renormalization 
function
${\cal Z}^{-1}(x,\mu^2)$ corresponding to a chiral symmetry breaking solution. 
 Note that the
mass  function $M(x)$ is of the same order as ${\cal Z}^{-1}(x,\mu^2)$ 
itself.  
Nevertheless,
the analytical chirally symmetric solution of Sec.~\protect\ref{sec: cp one}
(thin line) provides an excellent approximation (better than one part in
a thousand for $x > \mu^2$) for $x > M(x)$.
\label{fig: a(x)}}}
\vspace{0.5cm}
\end{figure}

We shall now examine dynamical chiral symmetry breaking for the $CP$
vertex in the absence of any explicit symmetry breaking by a nonzero
bare mass, as before.  Even for solutions exhibiting dynamical symmetry 
breaking, it is to be 
expected that the analytic result derived for
${\cal Z}(x)|_{M(x)=0}$ [Eq.~(\ref{eq: cp z analytic})] remains valid as long
as $x$ is large  compared to $(M(x)/\nu)^2$ and $\epsilon$ is
sufficiently small.  That this is indeed the case is illustrated in 
Fig.~\ref{fig: a(x)},
where we show a typical example of ${\cal Z}^{-1}(x)$ for a solution which breaks
chiral symmetry.  In this Figure, as well as in the rest of this
Section, we shall be dealing with the renormalized ${\cal Z}(x)$ and $M(x)$
instead of the unrenormalized quantities in the previous sections.  This makes
no essential difference to the physics of chiral symmetry breaking, although
it of course effects the absolute scale of ${\cal Z}(x)$. For a 
discussion
of the renormalization of the dimensionally regularized
theory we refer the reader to Ref.~\cite{dim_reg_2}.

\begin{figure}[htb]
  \centering
      \epsfig{angle=90,height=7.5cm,file=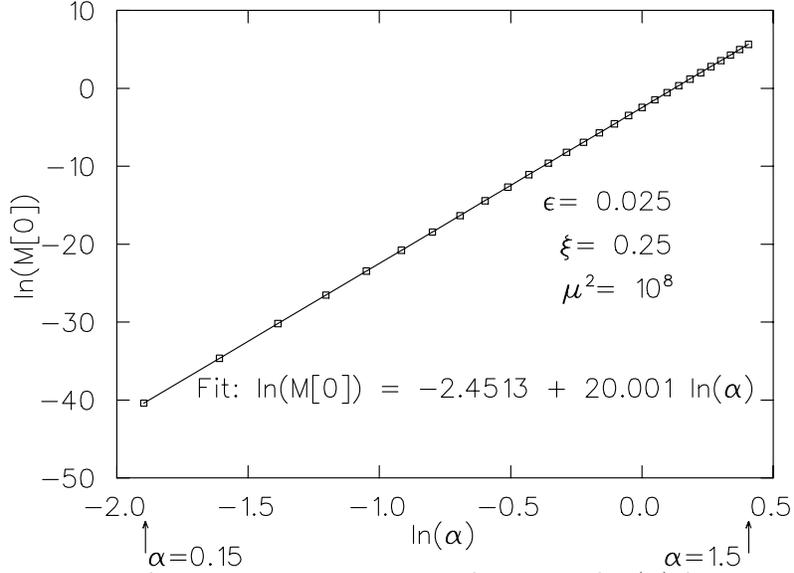}
  \parbox{130mm}{\caption{ The logarithm of the dynamical mass as a function
of ln($\alpha$) for $\epsilon=0.025$ (i.e. ${1 \over 2 \epsilon} = 20$).
The gauge parameter is fixed at $\xi = 0.25$ and the renormalization point is
$\mu^2 = 10^8$.  The open squares are the numerical values while the solid
line is a linear fit to these points. Note that the dependence on the
coupling expected in Eq.~(\protect\ref{eq: M(0) general form}) is reproduced
to high precision.
\label{fig: alpha_vs_mass}}}
\vspace{0.5cm}
\end{figure}

The comparison to the analytic result in Fig.~\ref{fig: a(x)} provides
a very convenient check on the numerics.  Another check is provided by
plotting the logarithm of $M(0)$ against the logarithm of the
coupling.  According to Eq.~(\protect\ref{eq: M(0) general form}) this
should be a straight line with gradient ${1 \over 2 \epsilon}$.  As
can be seen in Fig.~\ref{fig: alpha_vs_mass} not only does one
observe chiral symmetry breaking down to couplings as small as $\alpha=0.15$,
the expected linear behaviour is confirmed to quite high precision.

Although the numerics in $D < 4$ dimensions are clearly under control, the
extraction of the critical coupling (appropriate in four dimensions)
has proven to be numerically quite difficult. From the discussion in
Sections~\ref{sec_motivation} and~\ref{sec: rainbow}, we anticipate that
the logarithm of the dynamical mass has the general form
\begin{equation}
\ln\left({M(0) \over \nu}\right) \> = \> {1 \over 2 \epsilon} \> \ln\left({\alpha \over \alpha_c}\right)
\> + \> \ln\left(\overline M(0,\epsilon)\right)
\end{equation}
where the last term is subleading as compared to the first as $\epsilon$ tends
to zero.  For sufficiently small $\epsilon$, therefore, $\alpha_c$ is related
to the gradient of $\ln\left(M(0)\right)$ plotted against $\epsilon^{-1}$.

\begin{figure}[htb]
  \centering
      \epsfig{angle=90,height=8cm,file=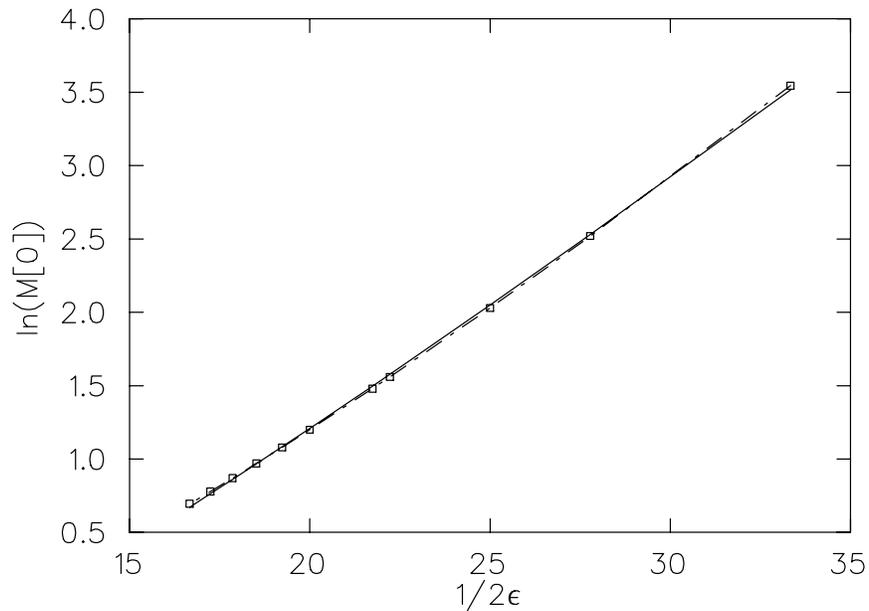}
  \parbox{130mm}{\caption{  The logarithm of the dynamical mass as a function
of ${1 \over 2 \epsilon}$ for a coupling of $\alpha = 1.20$. All other
parameters are as in Fig.~\protect\ref{fig: alpha_vs_mass}.  The open
squares are the numerical values while the two lines are fits (see main text)
\label{fig: logmass_vs_inv_epsilon}}}
\vspace{0.5cm}
\end{figure}

In Fig.~\ref{fig: logmass_vs_inv_epsilon} we attempt to extract $\alpha_c$ in this
way.  The logarithm of $M(0)$ was evaluated for $\epsilon$ ranging from 0.03
down to $\epsilon=0.015$ for a fixed gauge $\xi=0.25$.
The squares corresponds to a coupling constant $\alpha=1.2$, although some
of the points at lower $\epsilon$ have actually been calculated at smaller
$\alpha$ and then rescaled according to Eq.~(\protect\ref{eq: M(0) general form}).
At present we are unable, for these parameters, to decrease $\epsilon$
significantly further without a significant loss of numerical precision.
(We also note in passing that it is quite difficult numerically to move away
from small values of the gauge parameter; $\xi=20$, which, judging
from Fig.~\ref{fig: cut-off alpha_cr}, would not require a very high numerical
accuracy for $\alpha_c$, is unfortunately not an option.)

The two fits shown in Fig.~\ref{fig: logmass_vs_inv_epsilon}
correspond to two different assumptions for the functional form of
$\overline M(0,\epsilon)$, which is a priori unknown. The curves do
indeed appear to be well approximated by a straight line, however we
caution the reader that this does not allow an accurate
determination of $\alpha_c$ as the gradient is essentially determined
by the `trivial' dependence on $\log(\alpha)$ (more on this below). The solid line
corresponds to the assumption that the leading term in $\overline
M(0,\epsilon)$ has the same form as what we found in the rainbow
approximation, i.e.

\begin{equation}
\ln\left({M(0) \over \nu}\right) \> = \> {1 \over 2 \epsilon} \> \ln\left({\alpha \over \alpha_c}\right)
\> + \> c_1 \left({1 \over 2 \epsilon} \right)^{1 \over 3}\;\;\;.
\label{eq: fit1}
\end{equation}
With this form the fit parameters $\alpha_c$ and $c_1$ are found to be
\begin{equation}
\alpha_c \> = \> 0.966 \quad \quad c_1 \> = \> -1.15\;\;\;.
\end{equation}

Indeed, the critical coupling is similar to what is found in cut-off
based work (see Sec.~\ref{sec_motivation}; in Ref.~\cite{qed4_hrw} the
value was $0.9208$ for $\xi=0.25$).  At present it is difficult to
make a more precise statement, let alone differentiate between the two
curves plotted in Fig.~\ref{fig: cut-off alpha_cr}, as $\alpha_c$ is
quite strongly dependent on the functional form assumed in
Eq.~(\ref{eq: fit1}).  In fact, allowing an extra constant term on the
right hand side of Eq.~(\ref{eq: fit1}) reduces the critical coupling
to $0.920$ and the addition of yet a further term proportional to
$\epsilon^{1 \over 3}$ increases it again to $0.931$. As these numbers
appear to converge to something of the order of $0.92$ or $0.93$ one might think
that $\alpha_c$ has been determined to this precision.  However, it is not clear
that the functional form suggested by the rainbow approximation should be taken
quite this seriously.  The dashed line in  Fig.~\ref{fig: logmass_vs_inv_epsilon}
corresponds to a fit where the power of $\epsilon$ of the subleading term has
been left free, i.e.
\begin{equation}
\ln\left({M(0) \over \nu}\right) \> = \> {1 \over 2 \epsilon} \> \ln\left({\alpha \over \alpha_c}\right)
\> + \> c_1 \left({1 \over 2 \epsilon} \right)^{c_2}\;\;\;.
\label{eq: fit2}
\end{equation}
The optimum fit assuming this form for $\ln\left(M(0)\right)$ yields a power
quite different to ${1 \over 3}$ and a very much smaller $\alpha_c$:
\begin{equation}
\alpha_c \> = \> 0.825 \quad \quad c_1 \> = \> -0.801
\quad \quad c_2 \> = \> 0.688\;\;\;.
\end{equation}

To conclude this section, let us discuss why it is that
the functional form of the subleading term $\overline M(0,\epsilon)$
appears to be rather important even if $\epsilon$ is  already rather small.  The
reason for this is two-fold:  most importantly, although
the leading $\epsilon$ dependence of $\ln\left(M(0)\right)$ is indeed $\epsilon^{-1}$,
the coefficient of this term (leaving out the trivial $\alpha$ dependence)
is $\ln(\alpha_c)$.  As  $\alpha_c$ is rather close to 1 one therefore obtains
a strong suppression of this leading term, increasing the relative
importance of the subleading terms.  In addition, it appears as if the numerical
results favour a subleading term which is not as strongly suppressed (as a
function of $\epsilon$) as suggested by the rainbow approximation (i.e.
the power of $\epsilon^{-1}$ of the subleading term appears to be closer
to ${2 \over 3}$ rather than ${1 \over 3}$).  This again increases the
importance of the subleading terms.

\section{Conclusions and Outlook}
\label{sec: conclusion}

The primary purpose of this paper was to explore the phenomenon of
dynamical chiral symmetry breaking through the use of Dyson-Schwinger
equations with a regularization scheme which does not break the gauge
covariance of the theory, namely dimensional regularization. It is
necessary to do this as the cut-off based work leads to ambiguous
results for the critical coupling of the theory precisely because of
the lack of gauge covariance in those calculations. In particular,
this should be kept in mind when using the expected gauge invariance
of the critical coupling as a criterion for judging the suitability of
a particular vertex.

To begin with, we have shown on dimensional grounds alone and for an
arbitrary vertex, that in $D<4$ dimensions either a symmetry breaking
solution does not exist at all (in which case, however, it would also
not exist in $D=4$ dimensions) or it exists for all non zero values of
the coupling (in which case a chiral symmetry breaking solution exists
in $D=4$ for $\alpha > \alpha_c$).  For Dyson-Schwinger analyses
employing the rainbow and CP vertices we have shown that it is the
second of these possibilities which is realized.  For these symmetry
breaking solutions the limit to $D = 4$ is necessarily discontinuous
and so the extraction of the critical coupling of the theory (in 4
dimensions) is not as simple as in cut-off regularized work.

We next turned to an examination of symmetry breaking in the
rainbow approximation in Landau gauge, both analytically and numerically.
Indeed, for this vertex one could rewrite the (linearized)
Dyson-Schwinger equation as a Schr\"odinger equation in 4 dimensions
and appeal to standard results from elementary quantum mechanics to
explicitly show that the theory always breaks chiral symmetry if $D <
4$.  We also showed how the usual critical coupling $\alpha_c={\pi
\over 3}$ may be extracted from the dimensionally regularized work.

We concluded this work with an examination of the CP vertex.  By
making use of the gauge covariance of the theory we derived an exact
integral expression for the wavefunction renormalization function
${\cal Z}(p^2)$ of the chirally symmetric solution. Furthermore we 
obtained a compact expression
for this quantity which is an excellent approximation to the true
${\cal Z}(p^2)$ even for solutions which break the chiral symmetry.
Finally, we extracted the critical coupling corresponding to this
vertex and found that, within errors, it agrees with the standard
cut-off results.

\begin{figure}[htb]
  \centering
      \epsfig{angle=90,height=8cm,file=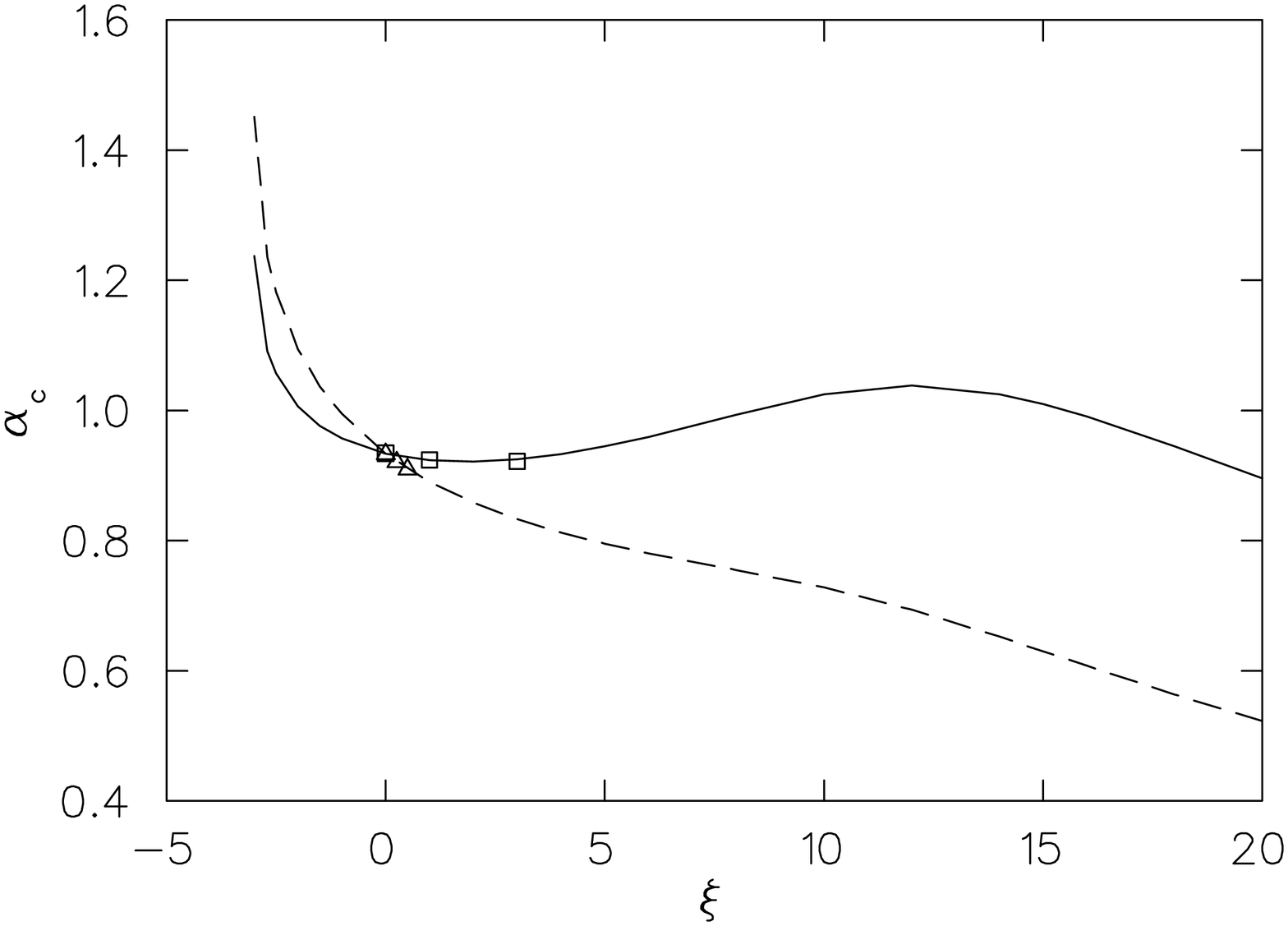}
  \parbox{130mm}{\caption{The critical coupling for the CP vertex.  The solid
line is taken from the bifurcation analysis carried out in
Ref.~\protect\cite{ABGPR}, which agrees with the numerical results
(open squares) of Ref.~\protect\cite{CPIV}.  The dashed line corresponds
to the bifurcation analysis carried out with the `gauge violating term'
removed (as suggested in Ref.~\protect\cite{dongroberts}) and
agrees with the numerical results (open triangles) of
Ref.~\protect\cite{qed4_hrw}.
  \label{fig: cut-off alpha_cr}}}
  \vspace{0.5cm}
\end{figure}

In the future, we plan to increase the numerical precision with which
we can extract this critical coupling for the CP vertex by an order of
magnitude or so.  The factor limiting the precision at present is that
when solving the propagator's Dyson-Schwinger equation with the CP
vertex by iteration the rate of convergence decreases dramatically as
$\epsilon$ is decreased below $\epsilon \approx 0.015$.  If this
increase in precision can be attained it will enable one to make a
meaningful comparison with the  cut-off based results shown
in Fig.~\ref{fig: cut-off alpha_cr}.

\begin{acknowledgements}
We would like to acknowledge illuminating discussions
with D.~Atkinson, A.\ K{\i}z{\i}lers\"{u}, V.~A.~Miransky and M.~Reenders. 
VPG is grateful to the members of the Institute for Theoretical
Physics of the University of Groningen for hospitality during his stay there.
This work was supported by a Swiss National Science Foundation
grant (Grant No. CEEC/NIS/96-98/7 051219), by the Foundation of Fundamental 
Research of 
Ministry of Sciences of Ukraine (Grant No 2.5.1/003) and by the
Australian Research
Council.

\end{acknowledgements}

\setcounter{equation}{0}
\makeatletter
\renewcommand\theequation{A\arabic{equation}}
\makeatother

\begin{appendix}
\begin{center}
{\bf APPENDIX A: CHIRAL SYMMETRY BREAKING IN RAINBOW APPROXIMATION}
\end{center}

In this appendix we show that the linearized version of Eq.~(\ref{masseq}),
i.e.
\begin{equation}
M(p^2)=(e \nu^{\epsilon})^2(3 - 2 \epsilon)\int \;\frac{d^Dk}{(2\pi)^D}\;
\frac{M(k^2)}
{k^2+m^2}\frac{1}{(p-k)^2}\;\;\;,
\label{masseq lin}
\end{equation}
has symmetry breaking solutions for all values of the coupling.
  Our aim here is to convert this equation to
a Schr\"odinger-like equation, which we do by introducing
the function
\begin{equation}
\psi(r)=\int\frac{d^Dk}{(2\pi)^D}\frac{e^{ikr}M(k^2)}{k^2+m^2}\;\;\;.
\end{equation}
With this definition we have
\begin{equation}
\left(-\Box + m^2\right)\psi(r)=\int\frac{d^Dk}{(2\pi)^D}e^{ikr}M(k^2)
\end{equation}
where $\Box$ is the  $D$-dimensional Laplacian and so
\begin{equation}
\left(-\Box +m^2\right)\psi(r)=e^2\nu^{2 \epsilon}(3 - 2 \epsilon)\int\frac{d^Dp}{(2\pi)^D}e^{ipr}
\int\frac{d^Dk}{(2\pi)^D}\frac{M(k^2)}
{k^2+m^2}\frac{1}{(p-k)^2}\;\;\;.
\end{equation}
 After shifting the integration variable $(p\rightarrow p+k)$
the last equation can be written in the form of a
Schr\"odinger-like equation
\begin{equation}
H\psi(r)=-m^2\psi(r),
\label{Schrodinger}
\end{equation}
where $H=-\Box + V(r)$ is the Hamiltonian, $E=-m^2$ plays the
role of an energy and the potential $V(r)$ given by
\begin{equation}
V(r)\>=\>-e^2\nu^{2 \epsilon}(3 - 2 \epsilon)\int\frac{d^Dp}{(2\pi)^D}\frac{e^{ipr}}{p^2}
\>=\>-\frac{\eta}{ r^{D-2}}\;\;\;,
\end{equation}
where
\begin{equation}
\eta \> = \> {\Gamma (1 - \epsilon ) \over 4 \pi^{2 - \epsilon}}
e^2\nu^{2 \epsilon}(3 - 2 \epsilon)\;\;\;.
\end{equation}
For $D=3$ the coefficient $\eta$ is $2 \nu \alpha$ while near $D=4$ it is ${3 \over \pi} \alpha \nu^{2 \epsilon}$.
It is well known from any standard course of quantum mechanics (see, for 
example, Ref.\cite{quant1}) that
potentials behaving as $1/r^{s}$ at infinity, with $s<2$, always support
bound states (actually, an infinite
number of them). In the present case this can be seen by considering the 
Schr\"odinger equation (\ref{Schrodinger}) for zero energy, i.e.  $E=0$. 
The $s$-symmetric wave function then satisfies the equation
\begin{equation}
\psi^{\prime\prime}+\frac{D-1}{r}\psi^\prime+\frac{\eta}{r^{D-2}}
\psi=0.  \end{equation} 
The solution finite at the origin $r=0$ is 
\begin{equation}
\psi(r)={\rm const.} \> \times \> r^{\epsilon - 1}J_{{1 \over
\epsilon}-1}\left(\frac{\sqrt{\eta}}{\epsilon}r^{\epsilon}\right).
\label{E0function}
\end{equation}
The Bessel function in (\ref{E0function}) has an infinite number of
zeros, which means that there is an infinite number of states with
$E<0$.

Returning now to Eq.~(\ref{Schrodinger}), we can estimate the lowest energy 
eigenvalue variationally by using 
\begin{equation}
\psi(r)=Ce^{-\kappa r}
\label{trialfunction}
\end{equation}
as a trial wavefunction.  Here $C$ is related to $\kappa$ by demanding that
$\psi$ is normalized, i.e.
\begin{equation}
|C|^2=\frac{(2\kappa)^D}{\Omega_D\Gamma(D)}\;\;\;,
\end{equation}
where $\Omega_D$ is the volume of a $D$-dimensional sphere.
Calculating the expectation value of the ``Hamiltonian'' $H$ on the
trial wave function in Eq.~(\ref{trialfunction}) we find
\begin{equation}
E_0(\kappa^2)=\langle\psi|H|\psi\rangle=\kappa^2
\left[ 1 - {2^{D-2} \over \Gamma(D)} \kappa^{D-4} \>  \eta \right]
\label{varenergy}
\end{equation}
The minimum of the ``ground state energy'' in Eq.~(\ref{varenergy}),
$E_0(\kappa)$, is reached at
\begin{equation}
\kappa^{4 - D} \> = \> (D-2)\> {2^{D-3} \over \Gamma(D)}\>   \eta 
\end{equation}
(for $D=3$ the parameter $\kappa$ is $\nu \alpha$ while near $D=4$ it is 
$\nu \left[{\alpha \over \pi/2}\right]^{{1 \over 2\epsilon}}$)
and is given by the expression
\begin{equation}
(E_0)_{\rm var}\>=\>-m^2
\>=\>\kappa^2 \left( 1 - {1 \over {D \over 2} - 1} \right) 
\> = \> \kappa^2 {D-4 \over D-2}\;\;\;,
\label{E_0}
\end{equation}
where the $1$ is the contribution from the kinetic
energy while the  $\left({D \over 2} - 1\right)^{-1}$
corresponds to the potential energy. For $D > 2$ the potential is attractive and
 for $2 < D < 4$ it 
is always larger than the kinetic energy, so for this case we get 
dynamical symmetry breaking for any value of $\alpha$.  For example, for 
$D=3$, one obtains $E_0=-\kappa^2=-\nu^2\alpha^2$ which coincides precisely 
with the  ground-state energy of the hydrogen atom (not surprisingly, as we 
have used the ground-state hydrogen wave function as our trial function). In
this case  the dynamical mass is $m = \nu\alpha$.

For $D$ near $4$, on the other hand, we obtain from Eq.~(\ref{E_0}) that 
\begin{equation}
m\simeq\nu({\epsilon})^{1/2}\left(\frac{\alpha}{\pi/2}\right)^
{1\over2 \epsilon}.
\end{equation}
This is of the general form anticipated in Section~\ref{sec_motivation},
 with $\alpha_c={\pi \over 2}$.
Indeed, for $D=4$,  the Schr\"odinger equation
(\ref{Schrodinger}) becomes an equation with the singular potential
\begin{equation}
V(r)=-\frac{\eta}{r^2},\quad \eta=\frac{\alpha}{\pi/3}.
\end{equation}
Again, it is known from standard  quantum mechanics\cite{quant2} that the spectrum
of bound states for such a potential depends on the
strength $\eta$ of the potential: it has an infinite number
of bound states with $E<0$ if $\eta>1$ and bound states are
absent if $\eta<1$. Thus, the true critical value for the coupling is
expected to be $\alpha_c=\pi/3$ instead of the $\alpha_c=\pi/2$
obtained with the help of the variational method (which made use of the
exponential Ansatz for the wavefunction and thus only gave an upper bound
for the energy eigenvalue). 

\end{appendix}

\setcounter{equation}{0}
\makeatletter
\renewcommand\theequation{B\arabic{equation}}
\makeatother

\begin{appendix}
\begin{center}
{\bf APPENDIX B: CHIRALLY SYMMETRIC QED FROM THE LANDAU-KHALATNIKOV TRANSFORMATION}
\end{center}
Because the CP vertex in the  chirally symmetric phase of
QED is gauge-covariant~\cite{dongroberts}
it is possible to derive an integral representation of the wavefunction
renormalization function ${\cal Z}(x)$ [see Eq.~(\ref{eq: massless a})] from the 
Landau-Khalatnikov transformation~\cite{LaKh}.  This transformation
relates the coordinate space propagator $\tilde S^\xi(u)$ in one gauge to the propagator 
in a different gauge.  Specifically, with covariant gauge fixing, we have
\begin{equation}
\tilde S^\xi(u) \> = \> e^{4 \pi \alpha \nu^{2 \epsilon} [\Delta(0) - \Delta(u)]}
\tilde S^{\xi=0}(u)
\end{equation}
where  $\Delta(u)$ is essentially the Fourier transform of the gauge-dependent
part of the photon propagator, i.e.
\begin{equation}
\Delta(u) \> = \>  -\xi \int {d^Dk \over (2 \pi)^D} 
{e^{-i k\cdot u} \over k^4}\;\;\;.
\end{equation}
Specifically, we obtain
\begin{equation}
\tilde S^\xi(u) \> = \> e^{-r (\nu\>  u)^{2 \epsilon}} \tilde S^{\xi=0}(u)
\label{eq: LK in D dim}
\end{equation}
where
\begin{equation}
r \> = \>- {\alpha \over 4 \pi} \> \Gamma(-\epsilon) (\pi)^\epsilon \> \xi\;\;.
\end{equation}
Substituting the coordinate-space propagator in Landau gauge, i.e.
\begin{eqnarray}
\tilde S^{\xi=0}(u) &=& \int {d^Dp \over (2 \pi)^D} {e^{i p \cdot u} \over {p \hspace{-5pt}/}} \nonumber \\
&=&{i \over 2\> \pi^{D/2}} \> \Gamma \left ({D \over 2}\right ) {{u \hspace{-5pt}/} \over u^D}
\;\;\;,
\end{eqnarray}
and  carrying out the inverse Fourier transform of Eq.~(\ref{eq: LK in D dim})
one obtains the wavefunction renormalization function in an arbitrary gauge,
namely
\begin{eqnarray}
{\cal Z}(x) &=& -{i \over 2\> \pi^{D/2}} \> \Gamma\left ({D \over 2}\right ) 
\int d^Du  e^{i p \cdot u} \>{p \cdot u \over u^D} 
e^{-r (\nu\>  u)^{2 \epsilon}} \nonumber \\
&=& x^{\epsilon \over 2} \> 2^{1 - \epsilon}  \> \Gamma(2-\epsilon)
\int_0^\infty  du\>  u^{\epsilon-1} e^{-r u^{2 \epsilon}} J_{2 - \epsilon}({\sqrt x} \> u)
\;\;\;.
\label{eq: res2}
\end{eqnarray}
Note that for small $x$ this function vanishes:
\begin{equation}
{\cal Z}(x) \>=\> { \Gamma \left({1 \over \epsilon}\right)
\over 4 \epsilon \> (2 - \epsilon)} \> r^{-{1 \over \epsilon}} \> x \> + \> 
O\left( x^2 \right)\;\;\;.
\end{equation}

It may be checked explicitly that
Eq.~(\ref{eq: res2}) is indeed a solution to 
Eq.~(\ref{eq: massless a}) for
arbitrary $D$ by making use of the expansion of Eq.~(\ref{eq: res2}) around
$x^{-\epsilon}=0$.  To be more precise, consider the RHS of
Eq.~(\ref{eq: massless a}) upon insertion of the power $y^\delta$ in the place
of ${\cal Z}(y)$.  Note that the integral converges only if
$\epsilon > \delta > \epsilon - 2$.
After some work the
result is that the R.H.S. of Eq.~(\ref{eq: massless a}) becomes
\begin{eqnarray}
1&+&\tilde c {2 - \epsilon \over 2} x^{\delta-\epsilon} \left [
-(1+\delta - \epsilon) {\Gamma(2-\epsilon) \over \Gamma(\epsilon)}
\sum_{n=-\infty}^{\infty} {\Gamma(\epsilon+n) \over \Gamma(2 - \epsilon+n)}
{1 \over n - \delta + \epsilon} \right ].
\label{insertdelta}
\end{eqnarray}
For $\epsilon < 1$ this may simplified further by applying
Dougall's formula (Eq. 1.4.1 in \cite{Bateman}) which, in this case,
reduces to
\begin{equation}
\sum_{n=-\infty}^{\infty} {\Gamma(\epsilon+n) \over \Gamma(2 - \epsilon+n)}
{1 \over n - \delta + \epsilon} \> = \> {\pi^2 \over \sin (\pi \epsilon)
\sin (\pi [\epsilon - \delta])}{1 \over \Gamma(1-\delta) \Gamma(2 + \delta -
2 \epsilon)}
\end{equation}
Using this result, Eq.~(\ref{insertdelta}) becomes
\begin{equation}
1 \> - \> \tilde c {2 - \epsilon \over 2} x^{\delta-\epsilon}
\frac{\Gamma(1-\epsilon)\Gamma(2-\epsilon)\Gamma(\epsilon-\delta)
\Gamma(2+\epsilon-\delta)} {\Gamma(2+\delta-2\epsilon)\Gamma(1-\delta)}.
\end{equation}
Note that, as opposed to the integral representation Eq.~(\ref{eq: massless a}),
this expression is defined for  $\delta$  outside the range 
$\epsilon > \delta > \epsilon - 2$ and so we may use it as an analytical
continuation of the integral.  Furthermore, note that this last expression
vanishes for integer $\delta\geq1$ hence we cannot obtain a simple power expansion 
around $x=0$ for ${\cal Z}(x)$ in this way.

On the other hand, an expansion in powers of $x^{-\epsilon}$ is possible.
If we seek a solution of the form
\begin{equation}
 {\cal Z}(x) = \sum\limits_{n=0}^{\infty} c_n x^{-n \epsilon}\;\;\;,
\label{series}
\end{equation}
we may equate the coefficients of equal powers of $x^{-\epsilon}$ after inserting
the series (\ref{series}) into both sides  of Eq.(\ref{eq: massless
a}).  This way we obtain the recurrence relation for
 the coefficients $c_n$ ($c_0=1$) as
\begin{equation}
{c_{n+1} \over c_n} = {\tilde c \over 2 (n+1)} \Gamma(-\epsilon)
\Gamma(3-\epsilon)
{\Gamma(1+\epsilon n+\epsilon) \Gamma(2 - \epsilon-\epsilon n) \over
\Gamma(2 - 2 \epsilon - \epsilon n) \Gamma(1+\epsilon n)}\;\;\;.
\label{eq:recursion1}
\end{equation}
This  may be solved leading
to
\begin{equation}
c_n=[{\tilde c\over2}\Gamma(-\epsilon)\Gamma(3-\epsilon)]^n
\frac{\Gamma(2-\epsilon)\Gamma(1+n\epsilon)}{\Gamma(2-\epsilon-
n\epsilon)n!}\;\;\;,
\end{equation}
so that finally we  obtain
\begin{equation}
{\cal Z}(x)=\Gamma(2-\epsilon) \sum\limits_{n=0}^\infty\frac
{\Gamma(1+n\epsilon)}{\Gamma(2-\epsilon- n\epsilon)n!}\left[{\tilde
c\Gamma(-\epsilon)\Gamma(3-\epsilon)\over2}x^{-\epsilon}\right]^n
\end{equation}
as the series expansion of the solution to
Eq.~(\ref{eq: massless a}).  The reader may check that this
coincides precisely with the corresponding expansion of
the solution obtained via the Landau-Khalatnikov transformations 
[Eq.~(\ref{eq: res2})].  The latter may be obtained by
changing the variable of integration from $u$ to $u/\sqrt x$,
expanding the exponential in the integrand and making use
of the standard integral
\begin{equation}
\int\limits_0^\infty x^{\alpha-1}J_\nu(cx)\>dx\>=\>2^{\alpha-1}c^{-\alpha}
\frac{\Gamma\left({\alpha+\nu\over 2}\right)}
{\Gamma\left (1+{\nu-\alpha\over 2}\right )}\;\;\;.
\end{equation}

\end{appendix}

\newpage


\end{document}